\documentclass[runningheads]{llncs}
\usepackage[T1]{fontenc}
\usepackage{graphicx}

\usepackage{amsmath}
\usepackage{amssymb}
\usepackage{graphicx}
\usepackage{subcaption}
\usepackage{booktabs}
\usepackage{algorithm}
\usepackage{algpseudocode}

\begin{document}
\title{
\large
Constella: A Novel Framework for Cost-Efficient Distributed AI Inference in LEO Space Data Centers}

\titlerunning{
Constella: Cost-Efficient Distributed AI Inference in Space Data Centers}
%


\author{Andrija Stanisic\inst{1}\orcidID{0009-0007-2120-1142} \thanks{Corresponding author.} \and
Milos Gravara\inst{1}\orcidID{0009-0006-7986-5243} \and
Juan Luis Herrera\inst{1}\orcidID{0000-0002-2280-2878} \and
Stefan Nastic\inst{1}\orcidID{0000-0003-0410-6315}
}
\authorrunning{A. Stanisic et al.}
%

\institute{
Distributed Systems Group, Technische Universität Wien, Austria \\
\email{\{a.stanisic,m.gravara,j.gonzalez,s.nastic\}@dsg.tuwien.ac.at}
}

\maketitle              

\vspace{-0.5cm}
\begin{abstract}

Space data centers built from Low-Earth Orbit (LEO) satellite constellations are gaining increasing attention as a scalable computing infrastructure. With access to abundant solar energy and high-throughput optical inter-satellite links, such constellations can run AI workloads directly in orbit, enabling new in-space application types while optimizing existing ones such as Earth observation. However, managing satellite constellations that combine heterogeneous satellite roles introduces a cost optimization challenge. Determining the appropriate constellation size and satellite role ratio for a given workload is challenging, as over-provisioning processing satellites increases system cost, while under-provisioning limits system efficiency. To enable cost-efficient execution of AI inference workloads in such space data centers, we present Constella, a novel framework that leverages DNN splitting for distributed AI inference in LEO satellite constellations. Constella comprises an offline resource identifier that determines the optimal ratio of processor-to-communicator satellites and an online assignment algorithm. The algorithm utilizes constellation telemetry to adaptively route data within the constellation and to ground stations. We evaluate Constella on a real-world satellite dataset across scenarios of increasing complexity. Results demonstrate a reduction in system cost by up to two orders of magnitude and lower end-to-end inference latency by up to 2.7$\times$ compared to other approaches, while maintaining no less than 81.9\% inference success rate.




\vspace{-0.3cm}
\end{abstract}

\vspace{-1cm}
\section{Introduction}
\label{section:introduction}


\vspace{-0.1cm}
Low-Earth Orbit (LEO) satellite constellations are emerging as a viable platform for building space data centers~\cite{GOOGLE,10.1145/3613424.3614271}. Equipped with solar arrays, optical inter-satellite links (ISLs), and compute accelerators, these constellations benefit from solar energy yields up to eight times higher than on Earth, while being capable of providing significant compute power~\cite{GOOGLE}. As artificial intelligence (AI) is widely regarded as a foundational general-purpose technology, the demand for computing infrastructure supporting AI workloads is projected to grow by orders of magnitude in the coming decades~\cite{10.1145/3613424.3614271}. This demand arguably exceeds what terrestrial energy sources alone can sustain, making space data centers a potential solution for provisioning AI compute at scale~\cite{GOOGLE}. Furthermore, LEO constellations are already pivotal for Earth Observation (EO) applications such as wildfire detection and disaster management~\cite{pusztai2024hyperdrive}. Traditional EO architectures follow a "bent pipe" design, where each satellite captures raw imagery and transmits it as-is to the ground station during intermittent communication windows~\cite{pusztai2024hyperdrive}. To reduce the volume of transmitted data, recent approaches opt for splitting the AI model~\cite{SLICE,COIN-LEO,FOOL}. This technique partitions the neural network at a specific boundary called the split layer. The satellite executes the early layers and transmits the resulting intermediate data to the ground station, which completes the tail-end inference~\cite{SLICE}. The choice of split layer thus directly balances on-board compute overhead against downlink data volume.

\vspace{-0.1cm}
However, traditional architectures concentrate both computational and communication loads on a single resource-constrained satellite. This is why existing constellations, such as SpaceDataHighway~\cite{agnew2012edrs}, introduce distinct satellite roles to distribute heavy workload and energy-expensive downlink transmission among different satellites. Satellites equipped with processing hardware can execute workloads on-board, but are more expensive to deploy and operate than their transmission-focused counterparts~\cite{monte1991mobile,denby2020orbital}. 
This cost asymmetry makes the ratio of processor to communicator satellites a key design decision. Since space data center infrastructure remains substantially more expensive than terrestrial alternatives~\cite{GOOGLE}, cost reduction is expected to contribute to wider adoption. For instance, over-provisioning processors inflates deployment cost, while under-provisioning communicators creates transmission bottlenecks. The problem is further complicated by the intermittent nature of LEO downlink. Due to high orbital velocities, direct ground contact is restricted to brief communication windows~\cite{ZHANG2025115}, and processed results must otherwise be buffered on-board~\cite{9554085}. Modern LEO constellations mitigate this limitation through high-throughput ISLs that reach up to 100~Gbps~\cite{pusztai2025stardust,bhattacherjee2019network} while consuming significantly less energy than downlink transmission~\cite{10930516}. This energy asymmetry enables a cost-efficient alternative where intermediate results are forwarded via ISL to satellites with imminent ground access, rather than requiring each satellite to independently perform energy-expensive downlink~\cite{wu2025enhancing}. Realizing this cost advantage requires addressing two research challenges, namely determining the minimum-cost subset of processor and communicator satellites from a given constellation that satisfies inference workload constraints, and coordinating their ISL communication and ground station transmission during execution.

\vspace{-0.1cm}
In this work, we propose Constella, a two-phase framework for cost-efficient distributed inference across heterogeneous LEO satellite constellations. First, in an offline phase, a constellation resource identifier jointly determines the Deep Neural Network (DNN) split layer and the minimum-cost constellation required to carry out the defined inference workload. This phase includes finding the cost-optimal number of processor and communicator satellites. Then, in an online phase, a latency-aware assignment algorithm dynamically assigns each processor to a communicator upon every inference task, leveraging ISL communication to minimize end-to-end delivery latency. Specifically, the main contributions of this work include:


\vspace{-0.2cm}
\begin{itemize}
    \item \textbf{Constella}: A novel two-phase framework for planning and executing cost-efficient distributed inference across heterogeneous LEO constellations. Constella separates satellites into processors and communicators, jointly optimizing constellation composition offline and processor-to-communicator assignment online, upon workload deployment. Evaluation on a real-world satellite dataset shows that Constella maintains at least 81.9\% inference success while reducing deployment cost by one to two orders of magnitude, mean end-to-end latency by up to 2.7$\times$, and total energy consumption by up to 74$\times$ across scenarios of varying complexity, compared to traditional baselines.



    \item \textbf{OCRI}: An \textbf{O}ffline \textbf{C}onstellation \textbf{R}esource \textbf{I}dentifier that determines the DNN split layer and the cost-minimal constellation configuration. OCRI, aimed at the offline phase of Constella, formulates a constrained optimization problem, which is then linearized into a mixed integer linear program and solved for global optimality. Across different scenarios, OCRI exhibits low execution overhead, with mean runtimes below 24\,ms per configuration.

    \item \textbf{LIA}: A \textbf{L}atency-aware \textbf{I}SL \textbf{A}ssignment algorithm for the online phase of Constella that minimizes end-to-end inference latency by dynamically assigning processors to communicators. LIA leverages gossip-based state dissemination to construct a set of eligible communicators based on buffer capacity, energy feasibility, and communication window availability, then selects the satellite with the earliest ground contact. Across scenarios, LIA remains lightweight, requiring less than 5\,ms of execution time per orbit.

\end{itemize}
\vspace{-0.3cm}
The remainder of this paper is organized as follows. Section~\ref{section:motivation} motivates the work and positions it within the existing literature. Section~\ref{section:constella} details considered system model and the Constella framework, focusing on its key underlying mechanisms. Experimental results are presented in Section~\ref{section:evaluation}, while Section~\ref{section:conclusion} concludes the work with future research directions.

\vspace{-0.6cm}
\section{Motivation \& Related Work}
\label{section:motivation}

\vspace{-0.3cm}
This section motivates the problem through an EO use case in Section~\ref{subsection:motivational_use_case}. Section~\ref{subsection:related_work} analyzes the state-of-the-art and positions our work within this context.

\vspace{-0.6cm}
\subsection{Motivational Use Case}
\label{subsection:motivational_use_case}

\vspace{-0.3cm}
\begin{figure}[h]
    \centering
    \includegraphics[width=0.6\linewidth]{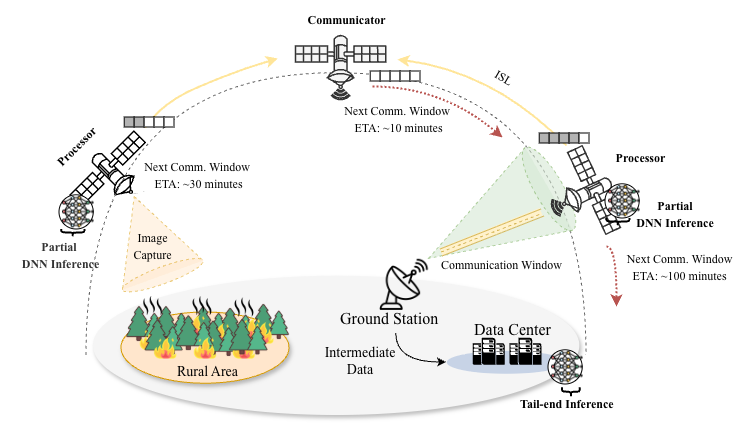}
    \caption{Motivational scenario for disaster detection.}
    \label{fig:scenario}
\end{figure}

\vspace{-0.3cm}

To better illustrate the motivation behind Constella, we present a use case on natural disaster prediction 
shown in Figure~\ref{fig:scenario}. This motivational use case serves as a showcase of the problems tackled by Constella in 
space data centers. To ensure the realism of the use case, the domain information is based on the proposal of Adeel \textit{et al.}~\cite{adeel2018survey}, and data for the technical and satellite aspects were extracted from real satellite constellations and DNN architectures~\cite{10.1145/1321400.1321404,tan2019efficientnet}.

Natural disasters such as wildfires, earthquakes, and flooding cause significant human and economic losses worldwide~\cite{united2025global}. These events result in significant annual fatalities and economic losses that frequently exceed hundreds of billions of USD~\cite{united2025global}. The primary objective in mitigating these impacts is early detection and prediction to ensure timely emergency responses~\cite{zschau2013early}. Traditional monitoring approaches address this by deploying ground-based sensor networks that continuously collect environmental data and propagate it into a terrestrial computing continuum for processing, analysis, and dissemination~\cite{adeel2018survey}. However, these approaches rely on the availability of reliable network connectivity. Many rural and remote regions lack such infrastructure, often due to the prohibitive cost of deploying and maintaining network equipment in areas with limited economic return~\cite{dlamini2021remote}. At the same time, rural regions are disproportionately affected by natural disasters~\cite{wang2022urban}. This creates a fundamental gap where monitoring and processing infrastructures are least available where they are needed the most. LEO satellite constellations offer an infrastructure-independent alternative. Equipped with imaging sensors and compute hardware, LEO satellites can capture imagery of affected regions, execute on-board inference, and transmit results to ground stations without relying on terrestrial networks~\cite{GOOGLE,SLICE}. This way, the satellites can achieve the same role as terrestrial approaches, providing continuous Earth observation and disaster management over areas that lack sufficient infrastructure. 

\vspace{-0.05cm}
For this motivational use case, consider a scenario in which governments of rural regions develop a DNN model for early wildfire detection and deploy it using the services of a LEO space data center. The data center offers two types of satellites at different price points, specifically processors and communicators. The processors capture Earth imagery, execute on-board inference, and perform downlink transmission, while the communicators are dedicated to downlink transmission at a lower cost. Both satellite types are equipped with high-throughput ISL terminals, enabling intra-orbit data propagation. Given these two satellite types, the first step is determining how many satellites of each type are required to serve the application. However, determining the right constellation dimensioning for a given workload is not trivial.
With $N$ satellites fulfilling one of the two previously defined roles and a DNN architecture comprising $|L|$ candidate split layers, the number of possible configurations is $2^N \cdot |L|$. For example, given the IRIDIUM~\cite{10.1145/1321400.1321404} constellation, which contains $N=66$ satellites and a DNN architecture such as ResNet50 with $|L|=182$ layers, the possible configurations reach approximately $2^{66}\cdot 182\approx1.35 \cdot 10^{22}$. For this relatively small constellation and simple neural network architecture, the search space is already intractable for manual exploration. Modern LEO constellations often comprise thousands of satellites~\cite{bhattacherjee2019network}, and state-of-the-art DNNs are becoming more complex~\cite{wang2024deepnet}, which further amplifies the combinatorial growth of the problem. Hence, an automated solution for cost-efficient constellation composition becomes necessary.

Despite optimal constellation composition, runtime operation remains challenging. Each processor must decide which communicator to forward intermediate data to. The goal is minimizing end-to-end latency, as it can be crucial for timely disaster response~\cite{zschau2013early}. Figure~\ref{fig:scenario} illustrates varying communication window timings among communicators. Some enable immediate downlink, while others require on-board buffering. Targeting the nearest communicator minimizes transfer delay but risks long buffering. Conversely, targeting the earliest ground contact risks overload. Processors can execute direct downlink transmission, bypassing communicators, but this process requires sufficient energy for both inference and transmission. This combination of objectives and dynamic conditions makes the assignment problem unsuitable for a static optimization.

\vspace{-0.5cm}

\subsection{Related Work}
\label{subsection:related_work}

Space data centers built from LEO satellite constellations have recently emerged as a recognized computing paradigm. Beals et al.~\cite{GOOGLE} present a vision for a future space AI infrastructure, projecting that terrestrial energy sources alone will be insufficient to sustain growing AI demand at scale. The problem of executing DNN inference across satellite-ground architectures has received considerable attention. Chen et al.~\cite{SLICE} propose SLICE, which jointly optimizes the DNN split layer and transmission scheduling to minimize energy consumption within a fixed satellite deployment. Yao et al.~\cite{LEOEdge} present LEOEdge, an inference platform that generates per-satellite models and applies layered scheduling to distribute tasks across a constellation. Xu et al.~\cite{COIN-LEO} propose COIN-LEO, which uses deep reinforcement learning to split a DNN into sequential submodels deployed across multiple homogeneous satellites for collaborative inference. Qiao et al.~\cite{10738397} formulate on-orbit distributed inference as a satellite-selection and data-partitioning problem, minimizing energy across a given constellation. Bleier et al.~\cite{10.1145/3613424.3614271} make a quantitative case for space microdatacenters. Authors define dedicated computational satellites that offload on-board processing from sensing ones, motivating role-specialized constellations. However, in all of these works, the satellite constellation is treated as a fixed input. The number of satellites and their roles are assumed rather than determined. Where the DNN split layer is optimized, it is done within a given constellation rather than jointly with it. As shown in Section~\ref{subsection:motivational_use_case}, the joint configuration space grows combinatorially with constellation size, making manual selection intractable. To the best of our knowledge, no existing work addresses this joint optimization under energy and deployment cost constraints, which represents the gap addressed in this work. 


Beyond constellation configuration, routing data within LEO constellations via ISLs to reduce transmission latency has also attracted recent research efforts. Jia et al.~\cite{jia2017collaborative} propose a collaborative scheme that offloads data among satellites via ISLs before ground contact, jointly scheduling offloading and downloading to maximize ground station throughput. Li et al.~\cite{li2024clhs} address the last-hop scheduling problem in large-scale constellations, proposing a centralized algorithm that issues download commands to satellites within the communication window of a ground station. In both works, routing decisions are computed centrally over a pre-planned schedule and do not operate at the granularity of individual inference tasks. To the best of our knowledge, no existing work addresses decentralized, per-task assignment of inference results from processor to communicator satellites under joint constraints on buffer capacity, energy feasibility, and communication window periods, which represents the gap addressed in this work.

\vspace{-0.6cm}
\section{Constella Framework Overview}
\label{section:constella}

\vspace{-0.3cm}


This section first introduces the system model and its underlying assumptions in Section~\ref{section:system_model}. It then presents the Constella framework and its two key components, OCRI and LIA, described in Section~\ref{subsection:ocri} and Section~\ref{subsection:lia}, respectively. OCRI executes on ground infrastructure prior to workload deployment and determines the minimum-cost constellation configuration that satisfies workload and environment constraints. The resulting configuration is then managed by LIA, which dynamically assigns processors to communicators during execution.

\vspace{-0.5cm}
\subsection{System Model} 
\label{section:system_model}

\vspace{-0.2cm}
For this model, let a LEO satellite constellation comprise $N\in\mathbb{N}_1$ satellites and a dedicated ground station. Drawing from the architectural decoupling established in constellations~\cite{agnew2012edrs,10.1145/3613424.3614271}, we propose partitioning the LEO constellation into two disjoint satellite roles. First, the processors, which are satellites equipped with imaging sensors, on-board compute hardware, ISL terminals, and antennas for downlink transmission. Second, the communicators, which carry ISL terminals and antennas that enable downlink transmission. Processors are more expensive than communicators  because they additionally carry dedicated compute accelerators~\cite{denby2020orbital}. Given this, we assume that, in the case of a LEO data center, the price for using processors, denoted as $\alpha\in\mathbb{R}^+$, is not lower than $\beta\in\mathbb{R}^+$, which denotes the price of usage of communicators. 
The LEO data center provides a set of processors, denoted by $\mathcal{X}$, and communicators, denoted by $\mathcal{Y}$, where $N = |\mathcal{X}| + |\mathcal{Y|}$. The numbers of activated processors and communicators are denoted by $X$ and $Y$, respectively, where $X \leq |\mathcal{X}|$ and $Y \leq |\mathcal{Y}|$.
Processing hardware efficiency is characterized by the energy cost of computation $p\in\mathbb{R}^+$ ($Wh/FLOP$) and the energy cost of transmission $q\in\mathbb{R}^+$ ($Wh/bit$), both of which are treated as a priori known parameters. Additionally, energy budgets per orbit for processors and communicators, denoted as $E^{proc}_{budget}, E^{comm}_{budget}$ respectively, and downlink channel throughput, denoted as $R^{max}$, are also considered to be known a priori. These assumptions reflect static hardware specifications and operational policies, known to the data center operator prior to deployment. We assume a homogeneous satellite environment, where all satellites of a given role share the same hardware efficiency parameters and initial energy budget.



Regarding the constellation topology, we consider full pairwise connectivity via multi-hop ISL routing~\cite{westphal_leo_2023}, with shortest-path routing treated as orthogonal to this work. We consider a static deployment snapshot where $X$ processors and $Y$ communicators are uniformly interleaved across the orbital plane. This inherently minimizes processor-to-communicator distances, while joint spatial role optimization remains promising future work, as acknowledged in Section~\ref{section:conclusion}. To enable decentralized assignment during execution, we assume a lightweight ISL-based gossip protocol. Because ISL transmissions incur negligible latency and energy costs compared to ground transmission~\cite{10930516}, communicators can periodically broadcast their current state without depleting limited resources. Specifically, each activated communicator $y \in \mathcal{Y}$ disseminates its buffer occupancy $buff(t)_y$, remaining energy budget $E^{comm}_{budget}(t)_y$, and time to next ground contact $t_y^g$. The gossip period is strictly shorter than $\Delta_t$, ensuring all processors maintain a consistent view of communicator states before each assignment decision.

Furthermore, the orbit period $\mathcal{T}$ comprises computation $\mathcal{T}^{comp}$, communication $\mathcal{T}^{comm}$, and idle $\mathcal{T}^{idle}$ phases, where $\mathcal{T} = \mathcal{T}^{comp} + \mathcal{T}^{comm} + \mathcal{T}^{idle}$. The $\mathcal{T}^{comp}$ corresponds to the orbital segment traversing an area of interest (AOI), that is, regions requiring critical monitoring such as earthquake-impacted zones. For simplicity, we consider a single shared AOI across all processors. During $\mathcal{T}^{comp}$, processors capture imagery and execute DNN inference at fixed intervals $\Delta_t$, yielding $I_{max} = \lfloor \mathcal{T}^{comp} / \Delta_t \rfloor$ potential tasks per orbit per processor. We assume that $\Delta_t$ is sufficient for full DNN execution on-board, which excludes inference latency from optimization. However, this remains a potential future research direction, as acknowledged in Section~\ref{section:conclusion}. Conversely, during $\mathcal{T}^{comm}$, processors or communicators transmit accumulated results to the ground station. To avoid the complexities associated with boundary-case data arrivals, any satellite entering $\mathcal{T}^{comm}$ is considered unavailable to receive additional data for downlink process, though it remains available as an ISL relay node. We assume that there is no overlap between $\mathcal{T}^{comm}$ and $\mathcal{T}^{comp}$, as the ground station is not considered to be part of the AOI (e.g., because it has access to Internet infrastructure). During $\mathcal{T}^{idle}$, satellites traverse orbital segments outside both the AOI and ground station visibility, performing neither inference nor downlink transmission.

Finally, the DNN model architecture is represented as an ordered set of layers $L = \{1, \ldots, |L|\}$. The split layer $l \in L$ defines the boundary of on-board execution. All layers from input up to and including layer $l$ are executed on the satellite. 
In case $l = |L|$, complete inference is executed on-board and only the final result is transmitted. The split layer induces two workload functions. The cumulative FLOPs, denoted with $W(l)\in\mathbb{R}^+$, needed to execute the first $l$ layers, and $D(l)\in\mathbb{R}^+$, the intermediate output size at layer $l$ in bits. The selection of $l$ therefore fundamentally balances computational cost against communication volume: larger values of $l$ increase $W(l)$ and on-board energy consumption, while reducing $D(l)$ and the resulting transmission time and energy.

\vspace{-0.4cm}
\subsection{OCRI: Offline Constellation Resource Identifier}
\label{subsection:ocri}

\vspace{-0.2cm}
The primary objective of the OCRI is to identify a minimum-cost constellation composition that serves a given inference workload under energy constraints. The workload is defined by a DNN architecture and a target number of inference tasks per orbit $I_{total}$. The optimization goal is to identify a triple $(l, X, Y)$ that minimizes deployment cost while complying with the energy constraints of each satellite. The resulting optimization problem is not linear due to the dependency of the workload functions $W(l)$ and $D(l)$ on the discrete split layer $l$. Since $W(l)$ and $D(l)$ are determined by the DNN architecture, they do not follow a closed-form expression over $l$ and cannot be treated as continuous variables. To linearize the problem, we introduce binary variables $z_l \in \{0,1\}$ for each candidate layer $l \in L$, with $\sum_{l \in L} z_l = 1$ ensuring that exactly one split layer is selected. The workload functions are then expressed as linear combinations $\sum_{l \in L} W(l) \cdot z_l$ and $\sum_{l \in L} D(l) \cdot z_l$, respectively, where $W(l)$ and $D(l)$ become known constants for each $l$. This yields the mixed-integer linear program (MILP) in Equation~\ref{eq:opt_milp}, jointly determining the split layer and the processor and communicator counts.


\begin{equation}
\scriptsize
\begin{aligned}
    \min_{X, Y, z} \quad 
    & \alpha \cdot X + \beta \cdot Y \\
    \mathrm{s.t.} \quad
    \quad \\ \mathrm{c_1}: \quad & (X + Y) \cdot R^{max} \cdot \mathcal{T}^{comm} \ge I_{total} \cdot \sum_{l \in L} D(l) \cdot z_l \\
    \mathrm{c_2}: \quad & X \cdot E^{proc}_{budget} \geq I_{total} \cdot \sum_{l \in L} W(l) \cdot z_l \cdot p \\
    \mathrm{c_3}: \quad & I_{total} \cdot (\sum_{l \in L} W(l) \cdot z_l \cdot p + \sum_{l \in L} D(l) \cdot z_l \cdot q) \leq X \cdot E^{proc}_{budget} + Y \cdot E^{comm}_{budget} \\
    \mathrm{c_4}: \quad & X \cdot I_{max} \geq I_{total} \\ 
    \mathrm{c_5}: \quad & X \in \mathbb{N}_1,\, Y \in \mathbb{N}_0,\, X\leq \mathcal{|X|}, Y \leq \mathcal{|Y|}, z_l \in \{0,1\}^{|L|}\ \land \sum_{l \in L} z_l = 1
\end{aligned}
\label{eq:opt_milp}
\end{equation}

\vspace{-0.15cm}


The objective minimizes total deployment cost as a weighted sum of processor and communicator counts. Constraint~$\mathrm{c_1}$ ensures that the aggregate downlink capacity covers the total intermediate data generated per orbit. Constraint~$\mathrm{c_2}$ guarantees that processors have sufficient energy for their share of inference tasks. Constraint~$\mathrm{c_3}$ bounds the total energy consumption by requiring that the combined computation ($p \cdot W(l)$) and communication ($q \cdot D(l)$) energy, summed across all $I_{total}$ tasks, does not exceed the aggregate energy budget of all satellites. Runtime load imbalance is handled by the online algorithm. 
Constraint~$\mathrm{c_4}$ bounds the per-processor task load by the maximum number of capture intervals within $\mathcal{T}^{comp}$. Constraint~$\mathrm{c_5}$ enforces integrality, requiring at least one processor while permitting constellations with no dedicated communicators, bounds processor and communicator counts to their respective pool sizes, and restricts the split point to $L$. Since OCRI executes offline on ground infrastructure with sufficient computational resources, solving the MILP to optimality is feasible, guaranteeing an optimal constellation configuration for the given workload.

\vspace{-0.45cm}
\subsection{LIA: Latency-Aware ISL Assignment}
\label{subsection:lia}

\vspace{-0.15cm}
LIA operates during the online stage of Constella, receiving the optimal configuration $(l, X, Y)$ from OCRI as input. The main objective of LIA is to assign processors to communicators in order to route each inference result to the ground station while minimizing end-to-end delivery latency and respecting energy constraints. LIA executes independently on each processor satellite at every capture interval $\Delta_t$, making a per-task routing decision after on-board inference completes. Algorithm~\ref{alg:lia} formalizes this procedure.

\begin{algorithm}[h]
\scriptsize
\caption{LIA}\label{alg:lia}
\begin{algorithmic}[1]
\Require $l, Y, R^{max}, q, E^{comm}_{budget}(t)_y, E^{proc}_{budget}(t), t^g_y, buff(t)_y, D(l), \mathcal{T}^{comm}$
\State $\mathcal{E} \gets \emptyset$
\ForAll{$y \in Y$}
    \If{$t^g_y = 0 \;\lor\; buff(t)_y + D(l) > R^{max} \cdot \mathcal{T}^{comm} \;\lor\; (buff(t)_y + D(l)) \cdot q > E^{comm}_{budget}(t)_y$}
        \State \textbf{continue}
    \EndIf
    \State $\mathcal{E} \gets \mathcal{E} \cup \{y\}$
\EndFor
\If{$\mathcal{E} \neq \emptyset$}
    \State $y^* \gets \arg\min_{y \in \mathcal{E}} \; t^g_y$
    \State Forward $D(l)$ to $y^*$ via ISL
\ElsIf{$E^{proc}_{budget}(t) \geq D(l) \cdot q$}
    \State Reserve energy $\rightarrow$ queue $D(l)$ for direct downlink
\Else
    \State Transmission infeasible
\EndIf
\end{algorithmic}
\end{algorithm}

At each interval, communicator states are available to all processors through the gossip protocol described in Section~\ref{section:system_model}. Given the received buffer occupancy, remaining energy budget, and estimated time to the next ground contact for each communicator, LIA first constructs a set of eligible communicators $\mathcal{E}$ (line~1). A communicator is excluded if it has already entered its communication window and can no longer accept additional data for the current pass, if the accumulated buffer including the new data exceeds the downlink capacity $R^{max} \cdot \mathcal{T}^{comm}$, or if the communicator lacks sufficient energy to transmit the resulting buffered data (line~3). Communicators that do not violate any of these conditions are added to $\mathcal{E}$ (line~6). If $\mathcal{E}$ is non-empty (line~8), the processor selects the communicator with the minimum estimated time to ground contact (line~9) and then forwards the result via ISL (line~10). If no communicator passes the eligibility checks, the processor falls back to direct downlink. However, there is a check if the processor has sufficient energy to transmit the data itself (line~11). In case it does, it reserves the required energy and stores the data for its own communication window (line~12). Otherwise, the transmission is infeasible (line~14). Since LIA operates without a central orchestrator, multiple processors may independently select the same communicator within the same interval. This can lead to transient buffer overloading, which can be resolved in subsequent intervals as updated states propagate through gossip protocol.


\vspace{-0.55cm}
\section{Evaluation}
\label{section:evaluation}

\vspace{-0.3cm}
This section presents a series of experiments as means to evaluate Constella. Section~\ref{subsection:experimental_setup} details the carried-out experiments, experimental frameworks, and evaluation objectives, while  Section~\ref{subsection:result_analysis} present the results.

\vspace{-0.5cm}
\subsection{Experimental setup}
\label{subsection:experimental_setup}

\vspace{-0.25cm}
To evaluate the performance of Constella, we compare it against two baseline approaches that represent prevalent design strategies. In the absence of existing approaches that address the same problem as Constella, we construct representative baselines that capture fundamental trade-offs between satellite resource planning and runtime communication assignment. Each baseline, like Constella, comprises an offline resource-identifying phase and an online assignment phase, enabling a fair comparison. The Naive Baseline (NB) utilizes all available satellites in the constellation ($X = |\mathcal{X}|, Y = |\mathcal{Y}|$) and selects the architectural midpoint of the DNN as the split layer ($l = \lfloor |L|/2 \rfloor + 1$), representing a heuristic that requires no optimization. During the online phase, each processor forwards its inference output to a fixed communicator determined by orbital position through a static, pre-computed assignment. The Traditional Baseline (TB) employs all available processing satellites ($X = |\mathcal{X}|, Y = 0$) and performs no on-board inference. Each satellite captures raw imagery and transmits it directly to the ground station during its communication window. This baseline effectively represents the traditional "bent pipe" architecture. Together, these two baselines span the design space from over-provisioned split inference to a communication-only pipeline, enabling assessment of decisions made by Constella.

\vspace{-1cm}
\begin{table*}[h]
\scriptsize
\centering
\caption{Evaluation scenarios and constellation configuration.}
\label{tab:configurations}

\begin{tabular*}{\textwidth}{@{\extracolsep{\fill}}lccccc}
\toprule
\multicolumn{6}{l}{\textit{Scenario Configuration}} \\
\midrule
Scenario & $|\mathcal{X}|$ & $|\mathcal{Y}|$ & $I_\text{total}$ & Model & $|L|$ \\
\midrule
extra-small & 8 & 2 & 100 & alexnet & 23 \\
small       & 40 & 10 & 500 & squeezenet1.0 & 67 \\
medium      & 400 & 100 & 1000 & resnet50 & 182 \\
large       & 4000 & 1000 & 10000 & swin-b & 311 \\
extra-large & 8000 & 2000 & 100000 & efficientnet-b0 & 329 \\
\bottomrule
\end{tabular*}
\begin{tabular*}{\textwidth}{@{\extracolsep{\fill}}lcc}
\multicolumn{3}{l}{\textit{Constellation Configuration}} \\
\midrule
Parameter(s) & Value(s) & Unit \\
\midrule
$R^\text{max}$ & 625\,000 & bps \\
$p$ & $1.63 \cdot 10^{-13}$ & Wh/FLOP \\
$q$ & $9.89 \cdot 10^{-8}$ & Wh/bit \\
$\alpha, \beta$ & $\alpha = 4.0, \ \beta = 1.0$ & -- \\
$E^{proc}_{budget}, E^{comm}_{budget}$ & $E^{proc}_{budget} = 1.5, E^{comm}_{budget} = 5.0$ & Wh \\
$\mathcal{T}^{comp}, \mathcal{T}^{comm}, \mathcal{T}^{idle}, \Delta_t$ & $\mathcal{T}^{comp} = 5100, \mathcal{T}^{comm} = 300, \ \mathcal{T}^{idle} = 900,\ \Delta_t = 10$ & s \\
\bottomrule
\end{tabular*}
\end{table*}

\vspace{-0.65cm}

The evaluation framework is implemented using the PyTorch
ecosystem for model management. The OCRI optimization problem is formulated and solved using the \texttt{python-mip}~\cite{python-mip} library. The \texttt{torchinfo}~\cite{torchinfo} library
is used to extract per-layer metrics, specifically the cumulative floating-point operations $W(l)$ and the intermediate output data size $D(l)$ for each candidate split layer $l\in L$. All models are analyzed using a fixed input tensor shape $(3, 224, 224)$, representing the standard input for the considered model architectures~\cite{10.1145/3065386}. To demonstrate the generality of the Constella, we evaluate it across five scenarios varying in constellation size, workload, and DNN architecture complexity. Following established data center cost models~\cite{10.1145/1921168.1921189}, we adopt a $\alpha{:}\beta = 4{:}1$ price ratio between processors and communicators, using the communicator cost as the base unit. Constellation parameters are derived from BUPT-1~\cite{10.1145/3636534.3649371} dataset. Given its 115~Wh battery and 25\% Depth of Discharge (DoD) limit, we allocate $E^{proc}_{budget}=1.5$~Wh and $E^{comm}_{budget}=5.0$~Wh to ensure orbital sustainability. Hardware efficiency metrics align with the NVIDIA Jetson Orin Nano~\cite{10.1145/3636534.3649371}. Table~\ref{tab:configurations} shows considered scenarios and constellation configurations. To ensure reproducibility, the complete framework implementation is made publicly available through Zenodo~\cite{stanisic_2026_20512270}.



The evaluation assesses Constella across four dimensions. First, the trade-off between system cost and inference success rate is examined to quantify the ability of Constella to identify minimal-cost configurations without sacrificing task completion. Second, end-to-end inference latency is compared across approaches to evaluate the effectiveness of adaptive routing over static assignment strategies. Third, total energy consumption per orbit is analyzed to assess the impact of constellation composition on the energy balance. Fourth, computational overhead of both offline and online phases is measured to confirm practical feasibility.

\vspace{-0.5cm}
\subsection{Result analysis}
\label{subsection:result_analysis}

\vspace{-0.25cm}

\begin{figure}[h]
\centering
\begin{minipage}[t]{0.49\textwidth}
    \centering
    \includegraphics[width=\linewidth]{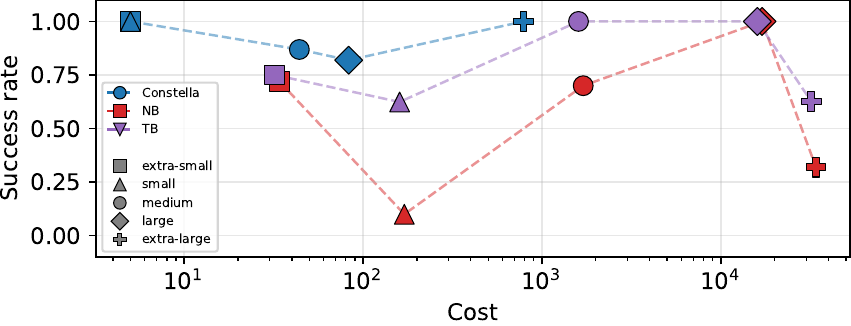}
    \caption{
    Deployment cost vs. success rate.}
    \label{fig:cost_success}
\end{minipage}
\hfill
\begin{minipage}[t]{0.49\textwidth}
    \centering
    \includegraphics[width=\linewidth]{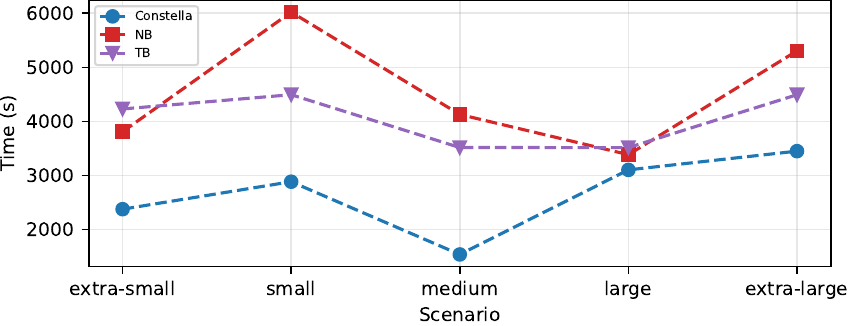}
    \caption{
    Mean inference latency.}
    \label{fig:latency}
\end{minipage}
\end{figure}

The key question for an identified constellation configurations is whether cost is minimized without sacrificing inference task completion. Figure~\ref{fig:cost_success} shows that Constella maintains at least 81.9\% success across all scenarios while reducing cost by one to two orders of magnitude compared to TB and NB. Success rate is defined as the fraction of generated images transmitted to the ground station within a single orbit. In extra-small and small scenarios, Constella achieves 100\% success at minimal cost, while NB degrades in the small scenario (10\% success) because its midpoint split layer produces an 8.96~Mbits intermediate tensor that overwhelms communicator buffers under static routing. In larger scenarios, Constella maintains high completion while requiring significantly lower cost. In the extra-large scenario, however, TB and NB degrade because their fixed processing and communication configurations exceed energy and buffer limits, whereas Constella adapts the split layer and constellation size to satisfy these constraints.

\vspace{-0.5cm}

\vspace{-0.2cm}

\vspace{-0.5cm}
Beyond cost, timely delivery of inference results is essential for latency-sensitive applications. Figure~\ref{fig:latency} shows that Constella reduces mean latency by 1.1--2.7$\times$ compared to TB and NB across scenarios. Mean latency is computed as the average time from image capture to completion of transmission to the ground station, with missed tasks assigned a one-orbit penalty of $\mathcal{T} = 6300~s$. In the medium scenario, Constella achieves the largest improvement, reaching 1541~s versus 4129~s for NB and 3518~s for TB. Its median latency of 991~s indicates that most tasks are delivered earlier, while a smaller tail waits for the next orbital pass. In the large scenario, latencies become more similar because Constella achieves its cost savings with only 3 communicators, which limits the routing advantage available over the more heavily provisioned baselines.


\vspace{-0.7cm}
\begin{figure}[h]
\centering
\begin{minipage}[t]{0.49\textwidth}
    \centering
    \includegraphics[width=\linewidth]{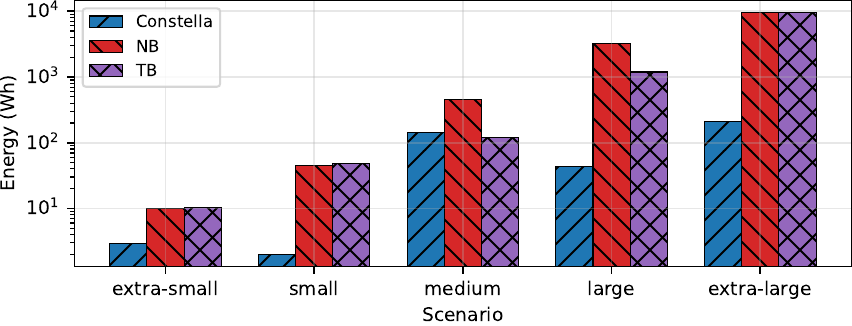}
    \caption{
    Energy consumption per orbit.}
    \label{fig:energy}
\end{minipage}
\hfill
\begin{minipage}[t]{0.49\textwidth}
    \centering
    \includegraphics[width=\linewidth]{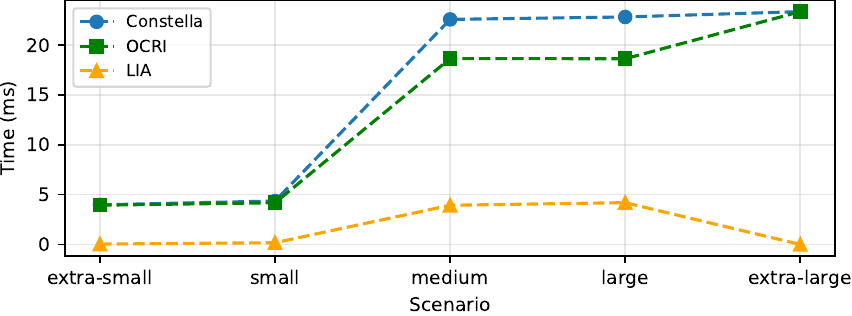}
    \caption{Mean execution time per orbit.}
    \label{fig:execution_time}
\end{minipage}
\end{figure}

\vspace{-0.7cm}

Reducing the number of active satellites also has direct implications for energy consumption. Figure~\ref{fig:energy} shows that Constella achieves the lowest total energy consumption in all scenarios except medium, consuming 3.5--45.5$\times$ less energy than TB and up to 74$\times$ less than NB. These savings arise from jointly selecting an efficient split point and smaller constellation configurations, which reduces both computation and communication energy. For example, in the large scenario, Constella uses only 23 satellites and consumes 43.0~Wh, compared to 3183.6~Wh for NB and 1191.0~Wh for TB. The only exception is the medium scenario, where TB consumes less total energy than Constella (119.1~Wh vs.\ 141.9~Wh) because it activates no communicators, while Constella expends additional communication energy to achieve substantially lower cost and latency.


Finally, we evaluate the execution overhead of OCRI and LIA to assess practical feasibility. Figure~\ref{fig:execution_time} shows mean execution times over 50 iterations. OCRI completes in 3.93--23.35~ms across all scenarios, demonstrating that the offline optimization remains efficient as problem complexity increases. The total LIA execution time shown in Figure~\ref{fig:execution_time} represents the cumulative time across all routing decisions per scenario, for one orbit, ranging from 0.04--4.19~ms. Dividing this range by the number of decisions per orbit yields a mean per-decision execution time of 0.50--4.04~$\mu s$, confirming negligible online overhead.
In the extra-large scenario, OCRI selects a configuration with no communicators ($Y=0$), so the LIA execution time is zero. Across all scenarios, the combined Constella execution time remains under 28~ms, confirming practical feasibility.


\vspace{-0.5cm}
\section{Conclusion}
\label{section:conclusion}

\vspace{-0.2cm}

Distributed inference in space data centers requires balancing deployment cost, energy consumption, and execution performance. This work introduced Constella, a two-phase framework for cost-efficient distributed inference in space. Constella jointly optimizes constellation configuration and split-layer selection offline, and performs adaptive processor-to-communicator assignment online. Evaluation shows that Constella maintains at least 81.9\% inference success while achieving one to two orders of magnitude lower deployment cost, up to 2.7$\times$ lower mean latency, and up to 74$\times$ lower total energy consumption than TB and NB across scenarios. 
The execution overhead of OCRI and LIA remains low, with mean runtimes below 24~ms per configuration and 5 ms per orbit for the cumulative execution of all assignment decisions, respectively, confirming practical feasibility. Future work will extend the Constella by modeling inference time as a variable to account for hardware heterogeneity. Additionally, we will investigate the joint optimization of satellite role placement within the constellation topology, including hybrid roles, to further improve system efficiency.


%
%
%

\vspace{-0.4cm}
\begin{credits}
\section*{Acknowledgements and Artifact Availability}
\vspace{-0.5cm}

This work was partly funded by the European Union under the Horizon Europe programme through the SNS JU (Grant Agreement No. 101192912, NexaSphere). Views expressed are those of the authors and do not necessarily reflect those of the EU or the SNS JU. \textbf{Disclosure of Interests.} The authors have no competing interests to declare that are relevant to the content of this article. The artifact is available in the Zenodo repository~\cite{stanisic_2026_20512270}.

\end{credits}

\bibliographystyle{splncs04}
\vspace{-0.5cm}

\bibliography{bibliography}

\end{document}